\documentclass[twocolumn,showpacs,preprintnumbers,amsmath,amssymb]{revtex4}  % double col.
\usepackage{graphicx}% Include figure files
\usepackage{dcolumn}% Align table columns on decimal point
\usepackage{bm}% bold math

\begin{document}

\title{Towards the perfect prediction of soccer matches}

% Place the author information here.  Please hand-code the contact
% information and notecalls; do *not* use \footnote commands.  Let the
% author contact information appear immediately below the author names
% as shown.  We would also prefer that you don't change the type-size
% settings shown here.

\date{\today}

\author{Andreas Heuer}
\affiliation{\frenchspacing Westf\"alische Wilhelms Universit\"at M\"unster, Institut f\"ur physikalische Chemie, Corrensstr.\ 30, 48149 M\"unster, Germany}
\affiliation{\frenchspacing Center of Nonlinear Science CeNoS, Westf\"alische Wilhelms Universit\"at M\"unster, Germany}
\author{Oliver Rubner}
\affiliation{\frenchspacing Westf\"alische Wilhelms Universit\"at
M\"unster, Institut f\"ur physikalische Chemie, Corrensstr.\ 30,
48149 M\"unster, Germany} \affiliation{\frenchspacing Center of
Nonlinear Science CeNoS, Westf\"alische Wilhelms Universit\"at
M\"unster, Germany}

%%%%%%%%%%%%%%%%% END OF PREAMBLE %%%%%%%%%%%%%%%%

\begin{abstract}

We present a systematic approach to the prediction of soccer
matches. First, we show that the information about chances for
goals is by far more informative than about the actual results.
Second, we present a multivariate regression approach and show how
the prediction quality increases with increasing information
content. This prediction quality can be explicitly expressed in
terms of just two parameters. Third, by disentangling the
systematic and random components of soccer matches we can identify
the optimum level of predictability. These concepts are exemplified
for the German Bundesliga.

\end{abstract}


%\baselineskip24pt


%\pacs{63.50.Lm,64.60.De,66.30.Dn}


\maketitle

\section{Introduction}

 One
important field is the prediction of soccer matches. In
literature different approaches can be found. In one type of
models \cite{Lee97,Dixon97,Dixon98,Rue00} appropriate parameters
are introduced to characterize the properties of individual teams
such as the offensive strength. Of course, the characterization of team strengths
is not only restricted to soccer; see, e.g., \cite{Sire}.  The specific values of these
parameters can be obtained via Monte-Carlo techniques. These
models can then be used for prediction purposes and allow one to
calculate probabilities for individual match results. A key
element of these approaches is the Poissonian nature of scoring
goals \cite{Maher82,janke1,janke2}. Beyond these goals-based prediction
properties also results-based models are used. Here the final
result (home win, draw, away win) is predicted from comparison of
the difference of the team strength parameters with some fixed
values \cite{Koning00}. The quality of both approaches has been
compared and no significant differences have been found
\cite{Goddard05}. Going beyond these approaches additional
covariates can be included. For example home and away
strengths are considered individually or the geographical distance
is taken into account \cite{Goddard05}. Recently, also the
ELO-based ratings have been used for the purpose of forecasting
soccer matches \cite{Hvattum}.

Recent studies suggest that statistical models are superior to lay and expert predictions but
have less predictive power than the bookmaker
odds \cite{Andersson,Song,Forrest,Hvattum}. This observation strongly suggests that either the information, used by the bookmakers,
is more powerful or, alternatively, the inference process, based on the same information, is more efficient. Probably, both
aspects may play a role.

When predicting soccer matches different key aspects have to be taken into account: (i) Choice of appropriate observables which contain optimum information about the individual team strengths, (ii) Definition and subsequent estimation of the team strength, (iii)Estimation of the outcome of a soccer match based on the two team strengths, (iv) Additional consideration of the stochastic (Poissonian) contributions to a soccer match. The final two aspects have been analyzed in detail in Ref.\cite{Heuer10}.

In the present work we concentrate on the first two aspects. Therefore we are restricting ourselves to predict the outcome of the second half of the season, i.e. summing over the final 17 matches (in the German Bundesliga). To reach this aim the stochastic aspects are somewhat easier to handle than for the prediction of a single match so that we can concentrate on (i) and (ii). However, all concepts can be also directly applied to the prediction of single soccer matches. Furthermore, our analysis can naturally be transferred to all other soccer leagues. As a key result we identify the level of optimum predictability and determine how close our actual inference approaches this optimum level.

It will turn out that the chances for goals are highly informative. They are provided by a professional sports journal (www.kicker.de) since the season 1995/96. In total we take into account all seasons until 2010/11. Since the definition of the chances for goals has slightly changed during the first years of the reporting period we have normalized the chances for goals such that their total number is identical in every season.

\section{Key elements of the prediction process}

\subsection{Systematic and stochastic effects in soccer matches}

Our general goal is the prediction of the future results of soccer matches. More specifically, we concentrate on the prediction of the outcome of the second half of the league tournament (German Bundesliga). This second half involves $N_2 = 17$ matches. We want to predict the final goal difference $\Delta G_2$ of each team after these $N_2$ match. A similar analysis could also be performed for points. We mention in passing that the information content of the goal difference about the team strength is somewhat superior to that of points \cite{Heuer09}.

In previous work we have defined the team strength $S_2$ of a team as the expected average goal difference when playing against all other 17 teams.  Strictly speaking, $S_2$ could be strictly determined if this team plays very often against the other 17 teams under identical conditions.

Let $\Delta G_2(N_2)$ denote the goal difference of some team after $N_2$ matches in the second half, normalized per match. Then $\Delta G_2(N_2)$ can be expressed as the sum of its strength $S_2$ and a random variable $\xi$, which denotes the non-predictable contributions in the considered matches.  In what follows we assume that the variance of $\xi$ is not correlated to the strength index $S_2$. Taking into account that the random contributions during different matches are uncorrelated one immediately obtains
\begin{equation}
\label{VG1}
Var(\Delta G_2(N_2)) = Var(S_2) + V_2/N_2
\end{equation}
where $V_2$ describes the variance of the random contribution during a single match and $Var(S_2)$ reflects the variance of the distribution of team strengths in the league \cite{Heuer09}. The $1/N_2$-scaling simply expresses that the statistical effects average out when taking into account a larger number of matches. This scaling only breaks down for $N_2$ close to unity because then the goal difference also depends on the strength of the opponent. In practice it turns out that for $N_2 > 4$ the difference of the $N_2$ opponents has sufficiently averaged out.  This dependence on the number of considered matches has been explicitly analyzed in Ref.\cite{Heuer09,heuer_buch}. For the present set of data we obtain $Var(S_2) = 0.21$ and $ V_2 = 2.95$.  Actually, $V_{2}$ is very close to the total number of goals per match (2.85). This expectation is compatible with the assumption of a Poissonian process.

\subsection{Prediction within one season}

In an initial step we use information from the first half of the season to predict the second half.  The independent variable in the first half is denoted as $Y$, the dependent variable in the second half as $Z$.
As the most simple approach we formulate the linear regression problem $Z = bY$. In what follows all variables fulfill the condition that the first moment of the variable, if averaged over all teams, is strictly zero. Generalization is, of course, straightforward. The regression problem requires the minimization of  $ \langle (Z-\hat{Z})^2\rangle$ with respect to $b$ where $\hat{Z}=bY$ is the explicit prediction of $Z$. Inserting the resulting value of $b_{opt}$ yields for this optimum quadratic variation
\begin{equation}
\label{chi2}
\chi^2(Y) = Var(Z) \left [1 -  [corr(Y,Z)]^2 \right ]
\end{equation}
where $Var(Z$) denotes the variance of the distribution of $Z$ and

\begin{equation}
 corr(Y,Z) = \frac{\langle YZ  \rangle }{ \sqrt{Var(Y) Var(Z)}}
\end{equation}

the Pearson correlation coefficient between the variables $Y$ and
$Z$. This relation has a simple intuitive interpretation. The
higher the correlation between the variables $Y$ and $Z$ the
better the predictability of $Z$ in terms of $Y$.

To be somewhat more general, we consider the case that exactly $N_1 (\le 17)$ matches in the first half of the season have been taken into account to define the independent variable $Y$. Whenever we want express the dependence on $N_1$ we use the terminology $Y(N_1)$. Without this explicit dependence we always refer to $N_1 = 17$. To reduce the statistical errors we always average over different random selections of $N_1$ matches from the first half of the season.

\subsection{Choice of observables}

A natural choice for the variable $Y$ is the goal difference  $\Delta G_1$ during the first half. We always assume that the results have been corrected for the average home advantage in that season. The quality of the prediction is captured by $corr(Y,Z)$; see Eq.\ref{chi2}. From the empirical data we obtain $corr(Y=\Delta G_1,Z=\Delta G_2)=0.56$.

Are there other observables $Y$ which allow one to increase $ corr(Y,\Delta G_2) $ significantly beyond the value of 0.56? The scoring of goals is the final step in a series of match events. One may thus expect that there exist other match characteristics  which are more informative about the team strength. A possible candidate is the number of chances for goals.  We denote the chances for goals as $C_\pm$ and the goals as $G_\pm$. The sign indicates whether it refers to the considered team (+) or the opponent of that team (-).

In a next step one can define the goal efficiencies $p_{\pm}$ via the relation

\begin{equation}
G_\pm = C_\pm \cdot p_{\pm}.
\end{equation}

Here, $p_{+}$ denotes the probability that the team is able to convert a chance for a goal into a real goal and $1 - p_{-}$ that the team manages to not concede a goal after a chance for a goal of the opponent. Averaging over all teams and seasons one obtains $\langle p_{\pm} \rangle = 0.24$. In analogy to  $\Delta G$ we will mainly consider  the difference $\Delta C = C_+ - C_-$ for prediction purposes.

If the goal efficiencies strongly vary from team to team in an a priori unknown way the chances for goals contain only very little information about the actual number of goals. If, however, the goal efficiencies are identical for all teams the chances for goals are more informative than the goals themselves. In Appendix I this general statement is rationalized for a simple model.

\begin{figure}[tb]
\centering\includegraphics[width=0.9\columnwidth]{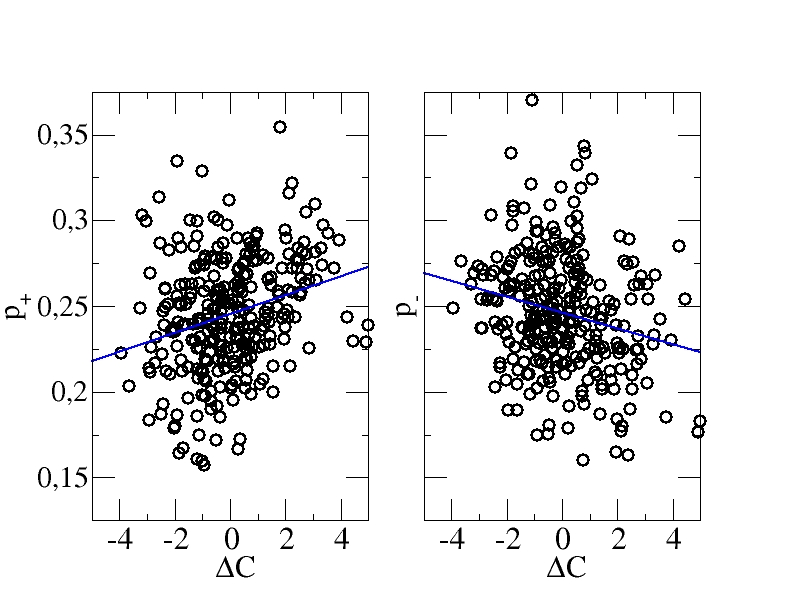}
\caption{The efficiency factors $p_\pm$ as a function of the differences of the chances for goals $\Delta C$ } \label{fig1}
\end{figure}

In Fig.\ref{fig1} the actual goal efficiencies $p_+$ after a season are shown together with the respective values of $\Delta C$. Naturally,  $\Delta C$ is strongly positively correlated with the team strength. Two effects are prominent.
(1) There is a slight correlation between $\Delta C$ and $p_+$. On average better teams have a slightly better efficiency to score goals. Analogous correlations exist between $p_-$ and $\Delta C$. (2) The goal efficiencies are widely distributed between approx. 15\% and 35\%. This observation would indicate that the information content of the chances for goals about the resulting team strength, defined in terms of scoring goals, is quite limited. Surprisingly, this is not true. For the correlation coefficient $corr(Y=\Delta C_1,Z=\Delta G_2)$ one obtains a value of 0.65 which is much larger than $corr(Y=\Delta G_1,Z=\Delta G_2)=0.56$.

To understand this high correlation for the chances for goals with the team strength we can discuss the reason for strong fluctuations of $p_{\pm}$ between the different teams. In general they are a superposition from two effects: (i) true differences between teams and (ii) statistical fluctuations, reflecting the random effects in the 34 soccer matches of the season. Both effects can be disentangled if one analyses the $N$-dependence of the variance of $p_{\pm}$. Whereas the statistical effects should average out for large $N$ the systematic effects remain for all $N$. In analogy to Eq.\ref{VG1} this can be written as
\begin{equation}
Var(p_{\pm}(N)) = Var(p_\pm) + const_\pm/N
\end{equation}

\begin{figure}[tb]
\centering\includegraphics[width=0.9\columnwidth]{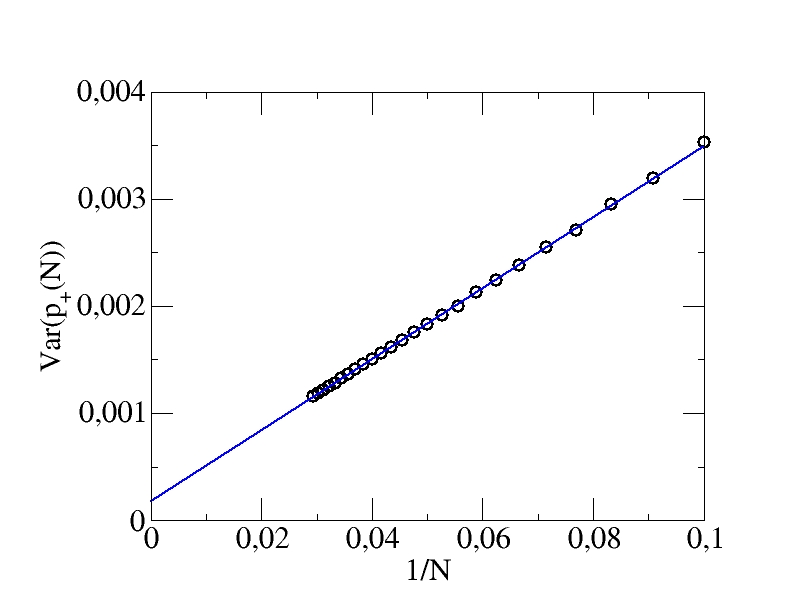}
\caption{ The variance of the distribution of goal efficiencies in dependence of the number of match days. } \label{fig2}
\end{figure}

$Var(p_\pm)$ can be interpreted as the true variance of the distribution of $p_\pm$ void of any random effects.  This N-dependence of $p_+(N)$ is explicitly shown in Fig.\ref{fig2}. Obviously, one obtains  very small values for $Var(p_+)$ and $Var(p_-)$ $(0.00017 \pm 0.00010$  and $0.00018 \pm 0.00010$, respectively). Thus, by far the largest contributions to the scatter of $Var(p_{\pm}(N=34))$ in Fig.\ref{fig1} is due to random effects. Stated differently, beyond the minor correlation between $p_{\pm}$ and $\Delta C$, shown in Fig.\ref{fig1}, the efficiency to score a goal out of a chance for a goal is basically the same for all teams!

To better understand the statistical properties of the chances for goals we again disentangle the systematic and random parts by writing
\begin{equation}
\label{VC1}
Var(\Delta C_1(N_1)) = Var(S_1) + \frac{V_1}{N_1}.
\end{equation}
One obtains $Var(S_1) = 2.66$ and $V_1 = 14.2$.  Based on this
relation it is possible to discuss the individual contributions to
the Pearson correlation coefficient $corr(\Delta C_1(N_1),\Delta
G_2)$. Using the independence of the random effects in the first
and the second half of the season one obtains
\begin{eqnarray}
\label{corryz}
& & corr(Y=\Delta C_1(N_1),Z=\Delta G_2) \nonumber \\
& = &  \frac{corr({S}_1,S_2)}{\sqrt{1+{V}_1/(N_1 Var({S}_1))}\sqrt{1+V_2/(17 Var(S_2))}}.
\end{eqnarray}

This expression clearly shows that there are three reasons why the prediction has intrinsic uncertainties, i.e. the correlation coefficient is smaller than unity. First, the team strength may change in the course of the season, i.e. $corr(S_1,S_2) < 1$. Since all parameters on the right side are explicitly known (see above) we can evaluate Eq.\ref{corryz}, e.g., for $N_1=17$. We obtain $corr(S_1,S_2) = 1.00$. Thus, the variation of the team strength during a single season is basically absent; see also Ref.\cite{Heuer10}. Second, the estimation of the team strength in the first half of the season is hampered by random effects, as expressed by $V_1/Var(S_1) > 0$. Of course, the larger the information content, i.e. the larger $N_1$, the better the prediction. For the chances for goals this ratio is given by $5.3$. If we had based $Y$ on $\Delta G$ rather than $\Delta C$   we would have obtained a value of 11.1. This comparison explicitly reveals why the chances for goals are more informative. Knowledge of the chances for goals of 10 matches is as informative as the goal differences of approx. 21 matches.  Third, the prediction of $\Delta G_2$ always has intrinsic uncertainties due to the unavoidable random effects in the second half of the season, i.e. $V_2/Var(S_2) > 0$.

Eq.\ref{corryz} allows one to define the limit of optimum prediction. It this case $Y$ would be explicitly given by $S_1$, i.e. $V_1 = 0$. This yields $corr(Y,Z=\Delta G_2) = 0.73$. This shows that the improvement of taking the chances for goals ($corr(Y=\Delta C_1,Z=\Delta G_2) = 0.65$) rather than the goals ($corr(Y=\Delta G_1,Z=\Delta G_2) = 0.56$) indeed is a major improvement relative to this optimum limit.

\subsection{Going beyond the present season}

Naturally, the prediction quality can be further improved by incorporating information from the previous season about the team strength. This additional variable is denoted as $X$. Here we consider the chances for goals of the previous season which we denote $ X=\Delta C_0$. One obtains $ corr(\Delta C_0, \Delta G_2) = 0.56$.  In principle one can again analyse the systematic and random contributions of $\Delta C_0(N_0)$. The corresponding $N_0$-dependent variance reads (see Eq.\ref{VC1})

\begin{equation}
Var(\Delta C_0(N_0)) = Var(S_0) + \frac{V_0}{N_0}
\end{equation}
with $Var(S_0) = 2.32$ and $V_0 = 14.1$. For reasons of comparison all relevant statistical parameters are summarized in Tab.\ref{tab1}.

\begin{table}
\label{tab1} \centering
  \begin{tabular}[t]{|c|c|c|}\hline
& $Var(S_i)$  & $V_i$ \\ \hline
      $i=0: \Delta C_0$  & 2.32 & 14.1\\ \hline
  $i=1: \Delta C_1$ & 2.66 &  14.2 \\\hline
 $i=2: \Delta G_2$ & $0.21$ & 2.95\\ \hline
       \end{tabular}
       \caption{ The different systematic and random contributions of the observables, relevant for this work.}
\end{table}

Of course, both values are close to $Var(S_1)$ and $V_1$. The small differences expresses the fact that the statistical properties of the first and the second half of the season are slightly different \cite{heuer_buch}.  Using the same reasoning as in the context of Eq.\ref{corryz} one finally obtains   $corr({S}_0,S_2)=0.88$ and $corr({S}_0,S_1)=0.86$. Both values are identical within statistical errors. This is compatible with the observation that the team strength does not vary within a season. The fact that both values are significantly smaller than unity shows, however, that there is a small but significant variation of the team strength between two seasons. For future purposes we use the average value of $corr({S}_0,S_{1,2})=0.87$ for the characterization of the correlation of the team strength between two seasons.

\section{Quality of the regression procedure}

\subsection{General information content}

\begin{figure}[tb]
\centering\includegraphics[width=0.9\columnwidth]{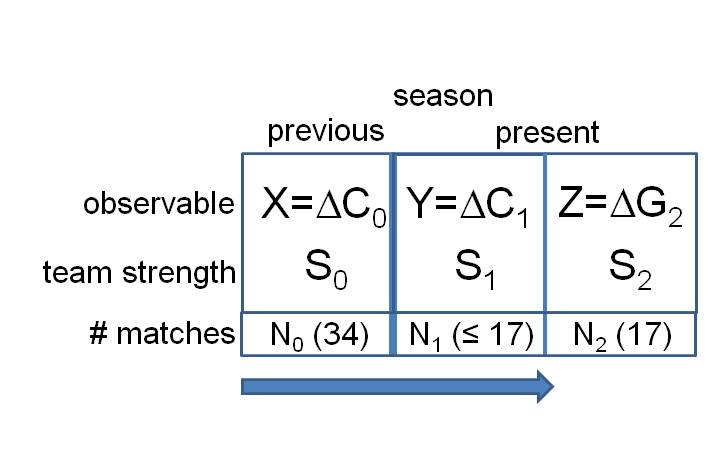}
\caption{Schematic representation of the general prediction setup.
} \label{fig3}
\end{figure}

For small $N_1$, i.e. at the beginning of the season, the information content about the strength of a team is quite limited. Therefore it is essential to incorporate also team information which is already available at the beginning of the tournament, i.e. reflects the  strength of this team from the past season. Thus, before the first match the prediction is fully based on $X$ and gradually with an increasing number of matches the variable $Y$ contains more and more information about the present team strength and thus will gain a stronger statistical weight in the inference process. This setup is sketched in Fig.\ref{fig3}. As discussed above we choose for $X$ the chances for goals of the previous season. The general relations, however, also hold beyond this specific choice.

Interestingly, the quality of the multivariate prediction can be expressed in analogy to  Eq.\ref{chi2} and reads
\begin{equation}
\label{chi2new}
\chi^2(X,Y) = \chi^2(Y) \left [1 - [corr(X-Y,Z-Y)]^2 \right ]
\end{equation}
where the partial correlation coefficient
\begin{equation}
c(X-Y,Z-Y) = \frac{corr(X,Z) - corr(X,Y)corr(Y,Z)}{\sqrt{1 - corr(X,Y)^2}\sqrt{1-corr(Y,Z)^2}}
\end{equation}
has been used. $\chi^2(Y)$ has been already defined in Eq.\ref{chi2}.
The second factor on the right-hand side of  Eq.\ref{chi2new} explicitly contains the additional information of the variable $X$ as compared to $Y$. One can easily show that in agreement with expectation Eq.\ref{chi2new} is completely symmetric in $X$ and $Y$.
Since Eq.\ref{chi2new} is non-standard it is explicitly derived in the Appendix II via some general arguments.

\subsection{Estimation of the team strength}

So far, we have identified $Z$ with the goal difference in the second half of the season which is composed of $S_2$ and the non-predictable random effects as expressed by $Var(\Delta G_2) = Var(S_2) + V_2/17$. Now we define
\begin{equation}
\label{chipract}
\tilde{\chi}^2(X,Y) = {\chi}^2(X,Y) - V_2/17.
\end{equation}

\begin{figure}[tb]
\centering\includegraphics[width=0.9\columnwidth]{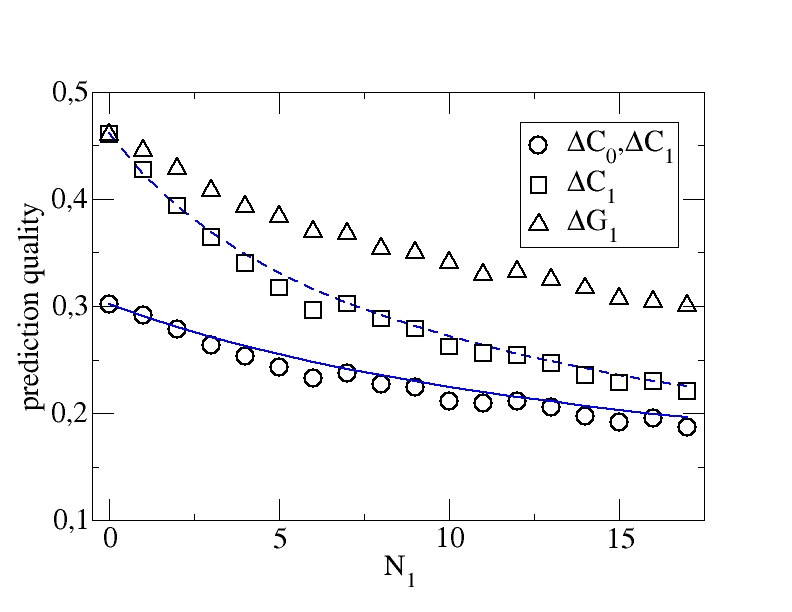}
\caption{The prediction quality of the team strength, determined
via $\sqrt{\tilde{\chi}^2(X,Y)}$, is shown as a function of the
number of  match days $N_1$. Different choices of variables are
shown. The solid lines are based on the explicit formulas for the
prediction quality.}\label{fig4}
\end{figure}

This can be interpreted as the statistical error for the prediction of the individual team strengths. In case of a perfect estimation of the team strengths one would have $\tilde{\chi}^2(X,Y)=0$. Mathematically this result can be derived by choosing $Z=S_2$ rather than $Z=\Delta G_2$ in Eq.\ref{chi2new}. After employing some straightforward algebraic manipulations of Eq.\ref{chi2new} one directly obtains $\tilde{\chi}^2(X,Y)$.

\section{Results}

\subsection{Numerical results}

For each value of $N_1$ we have performed a multivariate regression analysis, yielding $\chi^2(X,Y)$, and finally subtracted $V_2/17$. As before we have chosen several subsets of $N_1$ matches from the first half of the season to decrease the statistical error. Now we proceed in two steps. First, we neglect the contribution of $X$, i.e. the information from the previous season. The results are shown in Fig.\ref{fig4}.   One can see that (trivially) for $N_1=0$ the standard deviation in the estimation of the team strength is identical to the standard deviation of the $S_2$-distribution.  The longer the season, the more information is available to distinguish between stronger and weaker teams. Using the information of the complete first half of the season ($N_1=17$)
the statistical uncertainty decreases to 0.22. Here one can explicitly see the advantage of using the chances for goals rather than the goals themselves. Repeating the same analysis with the number of goals one would have an uncertainty of 0.30 after $N_1=17$ matches which is significantly higher than the value of 0.22, reported above. Second, when additionally incorporating the information from $X$, the statistical uncertainty is already quite small at the beginning of the season (0.3). Of course, during the course of the season it becomes even smaller. Even after 17 matches the additional gain of using $X$ is significant (0.22 vs. 0.19).

\subsection{Analytical results}

$\tilde{\chi}^2(X,Y)$ can be also calculated analytically by incorporating the statistical properties of the variables $X,Y$, and $Z$.  For future purposes we abbreviate $d=\tilde{V_1}/Var({S}_1)$. First, we have (using ($corr(S_1,S_2)=1$)

\begin{equation}
corr(Y=\Delta C_1(N_1),S_2) = \frac{1}{\sqrt{1+d/N_1}}.
\end{equation}

Furthermore, we express $corr(X,S_2)$ as
\begin{equation}
corr(X=\Delta C_0,S_2) = \frac{corr(S_0,S_{1,2})}{\sqrt{1+V_0/(17 Var(S_0))}} \equiv c,
\end{equation}

In analogy one obtains
\begin{equation}
corr(X=\Delta C_0,Y=\Delta C_1(N_1)) = \frac{c}{\sqrt{1+d/N_1}}.
\end{equation}

In summary, all information is contained in the two constants $c$ and $d$. A straightforward calculation yields $corr(X-Y,Z-Y) = c \sqrt{1-1/(1+d/N)}/\sqrt{1-c^2/(1+d/N)}$. Finally, one ends up with
\begin{equation}
\label{chitheo}
\tilde{\chi}^2(X,Y) = Var(S_2)\frac{\left ( 1-\frac{1}{1+d/N}\right )\left ( 1-c^2 \right )}{\left ( 1-\frac{c^2}{1+d/N}\right )}.
\end{equation}

Now we can compare the actual uncertainty, as already shown in Fig.\ref{fig4}, with the theoretical expectation, as expressed by the analytical result Eq.\ref{chitheo}. The results are included in Fig.\ref{fig4}. To reproduce the case without the variable $X$ one can simply choose $c=0$.  One can see a very close agreement with the actual data.

Is this good agreement to be expected? Actually, our analysis just
contains two approximations. First, we have chosen $corr(S_0,S_1)
=  corr(S_0,S_2)$ which, indeed, holds very well (see above).
Second, we have assumed that the team strength does not vary
during the first half of the season. As shown in
Ref.\cite{heuer_buch} the team strength fluctuates with a small
amplitude of approx. $A=0.17$ and with a decorrelation time of
approx. 7 matches. Since we average over different choices of
$N_1$ matches and, furthermore, restrict ourselves to the
prediction of the total second half, these temporal fluctuation
are to a large extent averaged out.

\section{Discussion}

The main goal was (i) to analyse the information content of
different observables and (ii) to better understand the limits of
the prediction of soccer matches. The prediction quality could be
grasped by the two parameters $c$ and $d$. One can easily see that
the theoretical expression for the prediction quality
Eq.\ref{chitheo} approaches the limit of perfect prediction in two
limits (i) For $c=1$ and $d=0$ the information from either the
previous or from the present season, respectively, perfectly
reflects the present team strength. (ii) For $ N_1 \rightarrow
\infty $ all random effects have averaged out so that only the
systematic effects remain.

This result can be easily generalized. For example one can  show
for the German Bundesliga that the market value, determined before
the season, is highly informative for the expected outcome. Taking
an appropriately chosen linear combination of different
observables one may slightly increase the value of $c$ but keeping
the general structure of Eq.\ref{chitheo} identical.

The same analysis could have been also performed by predicting
points rather than goal differences. Both observables are linearly
correlated via the simple relation $ P_2 = 0.61 S_2 + 23$.  In
analogy to $S_2$ the value of $P_2$ denotes the expected number of
points which a team gains in a match against an average team of
the league in a neutral stadium. Thus, an average team
($S-2=0$) on average gains 23 points per half-season.

\begin{figure}[tb]
\centering\includegraphics[width=0.9\columnwidth]{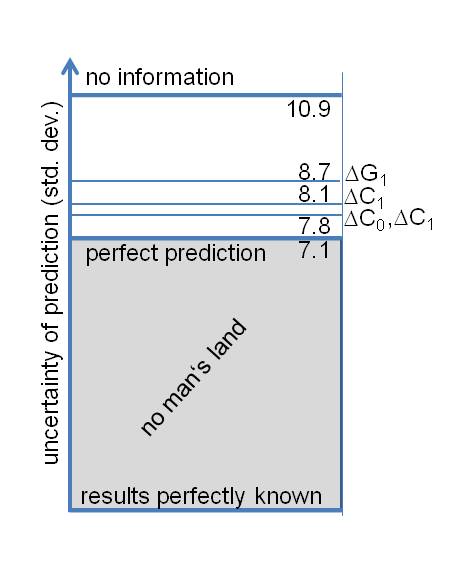}
\caption{The uncertainty of the prediction of the goal difference
of the second half when using the complete information of the
first half ($N_1 = 17$). Different choices of variables are shown.
Furthermore, the limit of perfect predictability is indicated. }
\label{fig5}
\end{figure}

One interesting question arises: is the residual statistical error
of $S_2$ for $N_1=17$ small or large? This question may be
discussed in two different scenarios. First, one may want to
predict the outcome of the second half of the league. In the
present context the uncertainty is given by $17\sqrt{\chi^2(X,Y)}
= 17\sqrt{\tilde{\chi}^2(X,Y)+V_2/17}$. These values are plotted
for different prediction scenarios in Fig.\ref{fig5}. One can see
how the additional information decreases the uncertainty of the
prediction. Most importantly, the {\it no man's land} below an
uncertainty of $\sqrt{17 V_2}=7.1$ cannot be reached by any type of prediction.
The art of approaching this perfect prediction thus resorts to
decrease the present value of 7.8 to a value closer to 7.1.
Second, one may be interested  in the prediction of a single
match. This case is somewhat different. Since the team
fluctuations are very difficult to predict the fluctuation
amplitude $A=0.17$ \cite{heuer_buch} serves as a scale for estimating the quality of
prediction. If the uncertainty is much smaller than $A$ any
further improvement would not help. In the present case the
statistical error is close to $A$ so that a further reduction of
$\tilde{\chi}^2(X,Y)$ would still be relevant for prediction
purposes.

Note that the chances for goals are not  completely objective
observable because finally also the subjective judgement of a
sports journalist may influence its estimates. In this sense the
high information content of chances for goals indicates that the
subjective component is quite small and the general definition is
very reasonable. Of course, in the future one may look for
strictly objective match observables taken by companies such as Opta and
Impire to further improve the information content. 

We gratefully acknowledge helpful discussions with D. Riedl, B. Strauss, and J. Smiatek.

\section{Appendix I}

Here we consider a simple example of a fictive coin-tossing tournament where the head appears with probability $p$ which in this simple example is given by 1/2. A team is allowed to toss the coin $M$ times per round.  In the first round this results in $g_{1}$ times tossing the head. Thus, in the first round one has observed the number of tosses $M$ as well as the number of heads $g_{1}$. In the relation to soccer $M$ would correspond to the number of chances for goals and $g_1$ to the number of goals in that match. In order to keep the argument simple we assume that $M$ is a constant whereas in a real soccer match $M$ can naturally vary. How to predict the expected number of goals $g_2$ in the next round? Here we consider two different approaches. (1) The prediction is based on the achievement of the first round, i.e. on the value of $g_1$. Then the best prediction is $g_2 = g_1$. The variance of the statistical error of the prediction can be simply written as $\sum_{g_1,g_2}p(g_1)p(g_2)(g_1-g_2)^2$ where $p(g)$ is the binomial distribution. A straightforward calculation yields for this variance a value of $2Mp(1-p)$. (2) The prediction is based on the knowledge of tossing attempts. The optimum prediction is, of course, $pM$. The variance of the statistical error is given by the binomial distribution, i.e. by $Mp(1-p)$. Stated differently, knowing the number of attempts to reach a specific goal (here tossing a head) is more informative that the actual number of successful outcomes as long as the probability $p$ is well known. 

\section{Appendix II}

Here we show a simple derivation of the chosen form of $\chi^2(X,Y)$. Let $d_{YZ}$ denote the solution of the regression problem $Z = dY$. Accordingly, $d_{YX}$ is the solution of the regression problem $X = dY$. In a next step one defines the new variables $\tilde{Z} = Z - d_{YZ}Y$ and $\tilde{X} = X - d_{YX}Y$. For these new variables the correlation with $Y$ is explicitly taken out. A straightforward calculation shows that the Pearson correlation coefficient $corr(\tilde{X},\tilde{Z})$ is exactly given by the partial correlation coefficient $corr(X-Y,Z-Y)$.

Now we consider the regression problem of interest $Z = aX + bY$. In a first step it is formally rewritten as
\begin{equation}
Z - d_{YZ} Y= a (X - d_{YX}Y) + (b - d_{YZ} + a d_{YX})Y.
\end{equation}
Using the above notation and introducing the new regression parameter $\tilde{b}$ we abbreviate this relation via
\begin{equation}
\tilde{Z} = a \tilde{X} + \tilde{b} Y.
\end{equation}
By construction the observable $Y$ is uncorrelated to $\tilde{X}$ and $\tilde{Z}$. Therefore the independent variable $Y$ does not play any role for the prediction of $\tilde{Z}$ so that effectively one just has a single-variable regression problem. Therefore one can immediately write
\begin{equation}
\chi^2(X,Y) = Var(\tilde{Z}) \left [ 1 - [corr(\tilde{X},\tilde{Z})]^2\right].
\end{equation}
The first factor is identical to $\chi^2(Y)$ whereas the Pearson correlation coefficient in the second factor is identical to $corr(X-Y,Z-Y)$. This concludes the derivation of $\chi^2(X,Y)$.

%\bibliography{doc}

\begin{thebibliography}{20}
\expandafter\ifx\csname natexlab\endcsname\relax\def\natexlab#1{#1}\fi
\expandafter\ifx\csname bibnamefont\endcsname\relax
  \def\bibnamefont#1{#1}\fi
\expandafter\ifx\csname bibfnamefont\endcsname\relax
  \def\bibfnamefont#1{#1}\fi
\expandafter\ifx\csname citenamefont\endcsname\relax
  \def\citenamefont#1{#1}\fi
\expandafter\ifx\csname url\endcsname\relax
  \def\url#1{\texttt{#1}}\fi
\expandafter\ifx\csname urlprefix\endcsname\relax\def\urlprefix{URL }\fi
\providecommand{\bibinfo}[2]{#2}
\providecommand{\eprint}[2][]{\url{#2}}

\bibitem[{\citenamefont{Gembris et~al.}(2002)\citenamefont{Gembris, Taylor, and
  Suter}}]{suter}
\bibinfo{author}{\bibfnamefont{D.}~\bibnamefont{Gembris}},
  \bibinfo{author}{\bibfnamefont{J.}~\bibnamefont{Taylor}}, \bibnamefont{and}
  \bibinfo{author}{\bibfnamefont{D.}~\bibnamefont{Suter}},
  \bibinfo{journal}{Nature} \textbf{\bibinfo{volume}{417}},
  \bibinfo{pages}{506} (\bibinfo{year}{2002}).

\bibitem[{\citenamefont{Ben-Naim et~al.}(2007)\citenamefont{Ben-Naim, Redner,
  and Vazquez}}]{ben1}
\bibinfo{author}{\bibfnamefont{E.}~\bibnamefont{Ben-Naim}},
  \bibinfo{author}{\bibfnamefont{S.}~\bibnamefont{Redner}}, \bibnamefont{and}
  \bibinfo{author}{\bibfnamefont{F.}~\bibnamefont{Vazquez}},
  \bibinfo{journal}{Europhys. Lett.} \textbf{\bibinfo{volume}{77}},
  \bibinfo{pages}{30005} (\bibinfo{year}{2007}).

\bibitem[{\citenamefont{Ben-Naim1 and Hengartner}(2007)}]{ben2}
\bibinfo{author}{\bibfnamefont{E.}~\bibnamefont{Ben-Naim1}} \bibnamefont{and}
  \bibinfo{author}{\bibfnamefont{N.~W.} \bibnamefont{Hengartner}},
  \bibinfo{journal}{Phys. Rev. E} \textbf{\bibinfo{volume}{76}},
  \bibinfo{pages}{026106} (\bibinfo{year}{2007}).

\bibitem[{\citenamefont{Bittner et~al.}(2007)\citenamefont{Bittner, Nussbaumer,
  Janke, and Weigel}}]{janke1}
\bibinfo{author}{\bibfnamefont{E.}~\bibnamefont{Bittner}},
  \bibinfo{author}{\bibfnamefont{A.}~\bibnamefont{Nussbaumer}},
  \bibinfo{author}{\bibfnamefont{W.}~\bibnamefont{Janke}}, \bibnamefont{and}
  \bibinfo{author}{\bibfnamefont{M.}~\bibnamefont{Weigel}},
  \bibinfo{journal}{Europhys. Lett.} \textbf{\bibinfo{volume}{78}},
  \bibinfo{pages}{58002} (\bibinfo{year}{2007}).

\bibitem[{\citenamefont{Bittner et~al.}(2009)\citenamefont{Bittner, Nussbaumer,
  Janke, and Weigel}}]{janke2}
\bibinfo{author}{\bibfnamefont{E.}~\bibnamefont{Bittner}},
  \bibinfo{author}{\bibfnamefont{A.}~\bibnamefont{Nussbaumer}},
  \bibinfo{author}{\bibfnamefont{W.}~\bibnamefont{Janke}}, \bibnamefont{and}
  \bibinfo{author}{\bibfnamefont{M.}~\bibnamefont{Weigel}},
  \bibinfo{journal}{Eur. Phys. J. B} \textbf{\bibinfo{volume}{67}},
  \bibinfo{pages}{459} (\bibinfo{year}{2009}).

\bibitem[{\citenamefont{Sire and Redner}(2009)}]{Sire}
\bibinfo{author}{\bibfnamefont{C.}~\bibnamefont{Sire}} \bibnamefont{and}
  \bibinfo{author}{\bibfnamefont{S.}~\bibnamefont{Redner}},
  \bibinfo{journal}{Eur. Phys. J. B} \textbf{\bibinfo{volume}{67}},
  \bibinfo{pages}{473} (\bibinfo{year}{2009}).

\bibitem[{\citenamefont{Lee}(1997)}]{Lee97}
\bibinfo{author}{\bibfnamefont{A.}~\bibnamefont{Lee}},
  \bibinfo{journal}{Chance} \textbf{\bibinfo{volume}{10}}, \bibinfo{pages}{15}
  (\bibinfo{year}{1997}).

\bibitem[{\citenamefont{Dixon and Coles}(1997)}]{Dixon97}
\bibinfo{author}{\bibfnamefont{M.}~\bibnamefont{Dixon}} \bibnamefont{and}
  \bibinfo{author}{\bibfnamefont{S.}~\bibnamefont{Coles}},
  \bibinfo{journal}{Appl. Statist.} \textbf{\bibinfo{volume}{46}},
  \bibinfo{pages}{265} (\bibinfo{year}{1997}).

\bibitem[{\citenamefont{Dixon and Robinson}(1998)}]{Dixon98}
\bibinfo{author}{\bibfnamefont{M.}~\bibnamefont{Dixon}} \bibnamefont{and}
  \bibinfo{author}{\bibfnamefont{M.}~\bibnamefont{Robinson}},
  \bibinfo{journal}{The Statistician} \textbf{\bibinfo{volume}{47}},
  \bibinfo{pages}{523} (\bibinfo{year}{1998}).

\bibitem[{\citenamefont{Rue and Salvesen}(2000)}]{Rue00}
\bibinfo{author}{\bibfnamefont{H.}~\bibnamefont{Rue}} \bibnamefont{and}
  \bibinfo{author}{\bibfnamefont{O.}~\bibnamefont{Salvesen}},
  \bibinfo{journal}{The Statistician} \textbf{\bibinfo{volume}{49}},
  \bibinfo{pages}{399} (\bibinfo{year}{2000}).

\bibitem[{\citenamefont{Maher}(1982)}]{Maher82}
\bibinfo{author}{\bibfnamefont{M.}~\bibnamefont{Maher}},
  \bibinfo{journal}{Statistica Neerlandica} \textbf{\bibinfo{volume}{36}},
  \bibinfo{pages}{109} (\bibinfo{year}{1982}).

\bibitem[{\citenamefont{Koning}(2000)}]{Koning00}
\bibinfo{author}{\bibfnamefont{R.}~\bibnamefont{Koning}}, \bibinfo{journal}{The
  Statistician} \textbf{\bibinfo{volume}{49}}, \bibinfo{pages}{419}
  (\bibinfo{year}{2000}).

\bibitem[{\citenamefont{Goddard}(2005)}]{Goddard05}
\bibinfo{author}{\bibfnamefont{J.}~\bibnamefont{Goddard}},
  \bibinfo{journal}{International Journal of Forecasting}
  \textbf{\bibinfo{volume}{23}}, \bibinfo{pages}{51} (\bibinfo{year}{2005}).

\bibitem[{\citenamefont{Hvattum and Arntzen}(2010)}]{Hvattum}
\bibinfo{author}{\bibfnamefont{L.~M.} \bibnamefont{Hvattum}} \bibnamefont{and}
  \bibinfo{author}{\bibfnamefont{H.}~\bibnamefont{Arntzen}},
  \bibinfo{journal}{International Journal of Forecasting}
  \textbf{\bibinfo{volume}{26}}, \bibinfo{pages}{460} (\bibinfo{year}{2010}).

\bibitem[{\citenamefont{P.~Andersson}(2005)}]{Andersson}
\bibinfo{author}{\bibfnamefont{M.~E.} \bibnamefont{P.~Andersson},
  \bibfnamefont{J.~Edman}}, \bibinfo{journal}{International Journal of
  Forecasting} \textbf{\bibinfo{volume}{21}}, \bibinfo{pages}{565}
  (\bibinfo{year}{2005}).

\bibitem[{\citenamefont{C.~Song and Stekler}(2005)}]{Song}
\bibinfo{author}{\bibfnamefont{B.~B.} \bibnamefont{C.~Song}} \bibnamefont{and}
  \bibinfo{author}{\bibfnamefont{H.}~\bibnamefont{Stekler}},
  \bibinfo{journal}{International Journal of Forecasting}
  \textbf{\bibinfo{volume}{23}}, \bibinfo{pages}{405} (\bibinfo{year}{2005}).

\bibitem[{\citenamefont{Forrest and Simmons}(2000)}]{Forrest}
\bibinfo{author}{\bibfnamefont{D.}~\bibnamefont{Forrest}} \bibnamefont{and}
  \bibinfo{author}{\bibfnamefont{R.}~\bibnamefont{Simmons}},
  \bibinfo{journal}{International Journal of Forecasting}
  \textbf{\bibinfo{volume}{16}}, \bibinfo{pages}{317} (\bibinfo{year}{2000}).

\bibitem[{\citenamefont{Heuer et~al.}(2010)\citenamefont{Heuer, Mueller, and
  Rubner}}]{Heuer10}
\bibinfo{author}{\bibfnamefont{A.}~\bibnamefont{Heuer}},
  \bibinfo{author}{\bibfnamefont{C.}~\bibnamefont{Mueller}}, \bibnamefont{and}
  \bibinfo{author}{\bibfnamefont{O.}~\bibnamefont{Rubner}},
  \bibinfo{journal}{Europhys. Lett.} \textbf{\bibinfo{volume}{89}},
  \bibinfo{pages}{38007} (\bibinfo{year}{2010}).

\bibitem[{\citenamefont{Heuer and Rubner}(2009)}]{Heuer09}
\bibinfo{author}{\bibfnamefont{A.}~\bibnamefont{Heuer}} \bibnamefont{and}
  \bibinfo{author}{\bibfnamefont{O.}~\bibnamefont{Rubner}},
  \bibinfo{journal}{Eur. Phys. J. B} \textbf{\bibinfo{volume}{67}},
  \bibinfo{pages}{445} (\bibinfo{year}{2009}).

\bibitem[{\citenamefont{Heuer}(2012)}]{heuer_buch}
\bibinfo{author}{\bibfnamefont{A.}~\bibnamefont{Heuer}},
  \emph{\bibinfo{title}{Der pefekte Tipp: Statistik des Fussballspiels}}
  (\bibinfo{publisher}{Wiley-VCH}, \bibinfo{year}{2012}).

\end{thebibliography}

\end{document}